\documentclass[aps,pre,twocolumn,superscriptaddress,10pt]{revtex4}
\usepackage[dvips]{graphics}
\usepackage{graphicx}
\usepackage{amsfonts}
\usepackage{amssymb}
\usepackage{amsmath}
\usepackage{subfigure}

\begin{document}

\title{\textbf{
Higher-order
effects and
ultra-short
solitons in left-handed metamaterials}}
\author{N.L. Tsitsas}
\affiliation{School of Applied Mathematical and Physical Sciences,
National Technical University of Athens, Zografos, Athens 15773, Greece
}
\author{N. Rompotis}
\affiliation{High Energy Physics Department, The Blackett Laboratory,
Imperial College,
London  SW7 2BW, UK
}
\author{I. Kourakis}
\affiliation{Centre for Plasma Physics, Queen's University Belfast
BT7 1 NN Northern Ireland, UK}
\author{P.G.\ Kevrekidis}
\affiliation{Department of Mathematics and Statistics, University of Massachusetts,
Amherst MA 01003-4515, USA}
\author{D.J.\ Frantzeskakis}
\affiliation{Department of Physics, University of Athens, Panepistimiopolis, Zografos,
Athens 15784, Greece }

\begin{abstract}
Starting from Maxwell's equations, we use
the reductive perturbation method to derive
a second-order and a third-order nonlinear Schr\"{o}dinger equation, describing
ultra-short solitons in nonlinear left-handed metamaterials. We find necessary conditions and
derive exact bright and dark soliton solutions of these equations for the electric and magnetic
field envelopes.
\end{abstract}

\maketitle



Electromagnetic (EM) properties of metamaterials with simultaneously
negative permittivity $\epsilon$ and permeability $\mu$ have
recently become a subject of intense research activity. Such
metamaterials
were experimentally realized recently in the microwave regime,
by means of periodic arrays of small metallic wires
and split-ring resonators (SRRs) \cite{exp}.
Many aspects of this class and other related types of metamaterials
have been investigated, and various potential applications have been
proposed \cite{reviews}.
So far, metamaterials have been mainly studied in the linear regime,
where $\epsilon$ and $\mu$ do not depend on the
EM field intensities. Nevertheless, {\it nonlinear} metamaterials, which may be created
by embedding an array of wires and SRRs into a nonlinear dielectric
\cite{zharov,agranovich,shadrivov}, may prove useful
in various applications. These include
``switching'' the material properties from left- to right-handed and back,
tunable structures with intensity-controlled transmission,
negative refraction photonic crystals, and so on.

EM wave propagation in nonlinear metamaterials can
be described by two coupled nonlinear Schr\"{o}dinger (NLS) equations for the
EM field envelopes \cite{lazarides-tsironis}.
Thus, bright-bright and dark-dark vector solitons of the Manakov type \cite{manakov} are
supported in the right-handed (RH) and left-handed (LH) regimes, respectively \cite{lazarides-tsironis}.
These findings paved the way for relevant studies, e.g., modulational instability \cite{kourakis}, and
bright-dark vector solitons \cite{shukla} in negative-index media.
%
Additionally, a scalar higher-order NLS (HNLS) equation was derived in \cite{scalora}
(assuming nonlinear response only in the electric properties of the metamaterial), and
was subsequently studied
\cite{wen-pre,shukla2}.
Coupled HNLS equations were also derived
\cite{wen-pra}, where
higher-order dispersion and nonlinear effects were included. However,
the relative importance of these effects was not studied in Ref. \cite{wen-pra},
although such an investigation should provide the necessary conditions for the formation
of few-cycle pulses in nonlinear metamaterials.

In this work, we present a systematic derivation of NLS and HNLS equations for the
EM field envelopes, as well as ultra-short solitons for left-handed (LH) metamaterials.
In particular, we use the reductive perturbation method \cite{rpm} to derive
from Faraday's and Amp\'{e}re's Laws a hierarchy of equations. Using such an approach,
i.e., directly analyzing Maxwell's equations,
we show that the electric field envelope is proportional to the magnetic field one (their ratio being the
linear wave-impedance). Thus, for each of the EM wave components we derive a {\it single} NLS
(for moderate pulse widths) or a {\it single} HNLS equation (for ultra-short pulse widths),
rather than a system of {\it coupled} NLS equations (as in Refs. \cite{lazarides-tsironis,kourakis,shukla,wen-pra}).
%
The
HNLS equation, which incorporates higher-order dispersive and nonlinear terms,
generalizes the one describing short pulse propagation in nonlinear optical fibers \cite{potasek,kodhas,potasek2,hasbook}.
Analyzing the NLS and HNLS equations, we find necessary conditions for the formation of
bright or dark solitons in the LH regime, and derive analytically approximate ultra-short solitons in
nonlinear metamaterials.

We consider lossless nonlinear metamaterials, characterized by
the effective permittivity and permeability \cite{zharov},
\begin{eqnarray}
\epsilon(\omega)&=&\epsilon_{0}\bigg(\epsilon_{D}(|\mathbf{E}|^2)-\frac{\omega_p^2}{\omega^2}\bigg),\quad \quad
\\
\mu(\omega)&=&\mu_{0}\bigg(1-\frac{F\omega^2}{\omega^2-\omega_{0NL}^2(|\mathbf{H}|^2)}\bigg),
\qquad  \label{eq:e-m}
\end{eqnarray}
where $\omega_p$ is the plasma
frequency, \emph{F} is the filling factor, $\omega_{0NL}$ is the
nonlinear resonant SRR frequency \cite{zharov}, while $\mathbf{E}$ and $\mathbf{H}$
are the electric and magnetic field intensities, respectively. In the linear limit,
$\epsilon_{D} \rightarrow 1$ and $\omega_{0NL} \rightarrow \omega_{\rm res}$ (where $\omega_{\rm res}$
is the linear resonant SRR frequency), and
LH behavior occurs in the frequency band
$\omega_{\rm res} < \omega < {\rm min}\{\omega_p , \omega_M \}$,
with $\omega_M = \omega_{\rm res} /\sqrt{1-F}$,
provided that $\omega_p > \omega_{\rm res}$.
On the other hand, a weakly
nonlinear behavior of the metamaterial can be approximated by
the decompositions \cite{lazarides-tsironis,kourakis,scalora,yskss}:
\begin{eqnarray}
\epsilon(\omega)&=&\epsilon_{L}(\omega)+\epsilon_{NL}(\omega;
|\mathbf{E}|^2),
\label{eq:eps} \\
\mu(\omega)&=&\mu_{L}(\omega)+\mu_{NL}(\omega; |\mathbf{H}|^2),
\label{eq:mu}
\end{eqnarray}
where $\epsilon_{L}=\epsilon_0 (1-\omega_p^2/\omega^2)$,
$\mu_{L} = \mu_0 [1-F\omega^2/(\omega^2-\omega_{\rm res}^2)]$,
while the nonlinear parts of the permittivity and permeability
are given by \cite{lazarides-tsironis,kourakis,scalora,yskss}:
$\epsilon_{NL}(|\mathbf{E}|^2) = \epsilon_0\alpha |\mathbf{E}|^2$,
and $\mu_{NL}(|\mathbf{H}|^2) = \mu_0\beta |\mathbf{H}|^2$;
here, $\alpha = \pm E_{c}^{-2}$ and $\beta$ are the Kerr
coefficients for the electric and magnetic fields, respectively,
$E_c$ being a characteristic electric field value.
The approximations (\ref{eq:eps})-(\ref{eq:mu}) are
physically justified considering that the slits of the SRRs are filled with a nonlinear dielectric
\cite{zharov,shadrivov}. Generally, both cases of focusing and defocusing dielectrics (corresponding,
respectively, to $\alpha>0$ and $\alpha<0$) are possible.
The magnetic Kerr coefficient $\beta$ can be found via the
dependence of $\mu$ on the magnetic field intensity \cite{zharov,shadrivov}.
Here, fixing $F=0.4$ and $\omega_p = 2\pi \times 10$ GHz, we will perform our analysis in
the frequency band from $2\pi \times 1.45$ GHz to $2\pi \times 1.87$ GHz,
considered also in Ref. \cite{lazarides-tsironis}. In this band,
SRRs are
LH media (with $\epsilon_L < 0$ and $\mu_L <0$ -- see Fig. 1),
$\alpha$ may
be either positive or negative, while $\beta$ is
positive \cite{lazarides-tsironis}.
Notice that, in principle, $\epsilon_{NL}$ and $\mu_{NL}$ may depend on {\it both} intensities
$|\mathbf{E}|^2$ {\it and} $|\mathbf{H}|^2$; such a case can also
be studied via the analytical approach we use below.

\begin{figure}[t]
\centering
\includegraphics[width=.38\textwidth, height=.2\textheight]{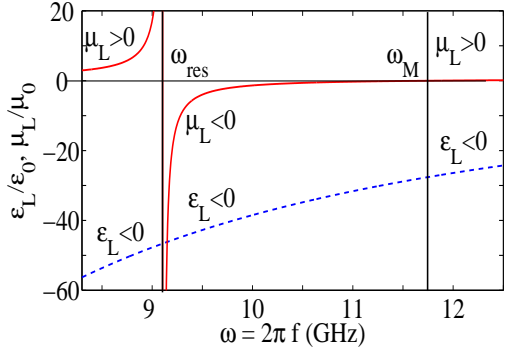}
\caption{
(Color online) The linear parts of the relative magnetic permeability, $\mu_L/\mu_0$ [solid (red) line],
and the electric permittivity, $\epsilon_L/\epsilon_0$ [dashed (blue) line] as functions of frequency,
for $F=0.4$ and $\omega_p = 2\pi \times 10$ GHz.
}
\label{fig1}
\vskip-0.35cm
\end{figure}
We consider the propagation along the $+\hat{\mathbf{z}}$
direction of a $x$- ($y$-) polarized electric (magnetic) field,
namely, $\mathbf{E}(z,t)=\hat{\mathbf{x}} E(z,t)$ and
$\mathbf{H}(z,t)=\hat{\mathbf{y}}H(z,t)$. Then, using the constitutive
relations (in frequency domain) $\mathbf{D}$=$\epsilon \mathbf{E}$ and $\mathbf{B}$=$\mu \mathbf{H}$
($\mathbf{D}$ and $\mathbf{B}$ are the electric flux density and the magnetic induction),
Faraday's and Amp\'{e}re's Laws respectively read (in the time domain):
%
\begin{align}
\partial_z E=-\partial_t(\mu\ast H),
\,\,\,\,\,
\partial_z H=-\partial_t(\epsilon \ast E),
\label{eq:Faraday-Ampere-laws}
\end{align}
where $\ast$ denotes the convolution integral, i.e., $f(t)\ast
g(t)=\int_{-\infty}^{+\infty}f(\tau)g(t-\tau)d\tau$.
%
Note that Eqs. (\ref{eq:Faraday-Ampere-laws})
may be used in either the RH or the LH regime:
once the dispersion relation $k_0 = k_0 (\omega_0)$
(for the wavenumber $k_0$ and frequency $\omega_0$) and the
evolution equations for the fields $E$ and $H$ are found,
then $k_0 > 0$ ($k_0 <0$) corresponds to the RH (LH) regime.
Alternatively, for fixed $k_0 >0$, one should shift the fields as $[E, H]^T
\rightarrow [\pm E, \mp H]^T$,
thus inverting the orientation of the magnetic
field and associated Poynting vector.
Here, we will assume that the wavenumber $k_0$ 
[see Eq. (\ref{eq:disp-rel}) below] will be $k_0 <0$ for the LH regime.

Now, we consider that the fields are expressed as
$[E(z,t), H(z,t)]^T = [q(z,t), p(z,t)]^T \exp[i(k_0 z-\omega_0 t)]$,
where
$q$ and $p$ are unknown field envelopes.
Nonlinear evolution equations for the latter can be found by
the reductive perturbation method \cite{rpm} as follows.
First, we assume that the temporal spectral  width of the
nonlinear term with respect to that
of the
quasi-plane-wave dispersion relation is characterized by the
small parameter $\varepsilon$ \cite{potasek,kodhas,potasek2,hasbook}.
Then, we introduce the slow
variables:
\begin{equation}
\label{eq:slow-variables} Z=\varepsilon^2z, \quad \quad
T=\varepsilon(t-k_{0}' z),
\end{equation}
where
$k_{0}' \equiv v_g^{-1}$ is the inverse of the group velocity (hereafter,
primes will denote derivatives with respect to $\omega_0$).
Additionallly, we express $q$ and $p$ as asymptotic expansions in terms of the parameter $\varepsilon$,
\begin{align}
q(Z,T)=&q_0(Z,T)+\varepsilon q_1(Z,T)+ \varepsilon^2 q_2(Z,T)
+ \cdots,
\label{eq:q-asympt}
\\
p(Z,T)=&p_0(Z,T)+\varepsilon p_1(Z,T)+ \varepsilon^2 p_2(Z,T)
+\cdots,
\label{eq:p-asympt}
\end{align}
and assume that the Kerr coefficients $\alpha$ and $\beta$
are of order $\mathcal{O}(\varepsilon^2)$ (see, e.g., \cite{lazarides-tsironis,potasek,kodhas}).
Substituting Eqs. (\ref{eq:q-asympt})-(\ref{eq:p-asympt}) into
Eqs. (\ref{eq:Faraday-Ampere-laws}), using Eqs. (\ref{eq:eps}),
(\ref{eq:mu}), and (\ref{eq:slow-variables}), and Taylor expanding
the functions $\epsilon_{L}$, and $\mu_{L}$, we arrive at the
following equations at various orders of $\varepsilon$:
\begin{eqnarray}
\mathcal{O}(\varepsilon^0): \,\, \mathbf{W}\mathbf{x}_0&=&\mathbf{0},
\label{eq:zero-order} \\
\mathcal{O}(\varepsilon^1): \,\, \mathbf{W}\mathbf{x}_1&=&-i\mathbf{W}'\partial_T\mathbf{x}_0,
\label{eq:first-order} \\
\nonumber
\mathcal{O}(\varepsilon^2): \,\,\mathbf{W}\mathbf{x}_2&=&-i\mathbf{W}'\partial_T\mathbf{x}_1
+\frac{1}{2}\mathbf{W}''\partial^2_T\mathbf{x}_0
\\ \label{eq:second-order}
&+&\frac{1}{2}k_0''\partial^2_T\mathbf{x}_0-i\partial_Z\mathbf{x}_0-\mathbf{A}\mathbf{x}_0,
\\ \nonumber
\mathcal{O}(\varepsilon^3): \,\,\mathbf{W}\mathbf{x}_3&=&-i\mathbf{W}'\partial_T\mathbf{x}_2
+\frac{1}{2}\mathbf{W}''\partial^2_T\mathbf{x}_1+\frac{i}{6}\mathbf{W}'''\partial^3_T\mathbf{x}_0
\\ \nonumber
&+&\frac{i}{6}k_0'''\partial^3_T\mathbf{x}_0
+\frac{1}{2}k_0''\partial^2_T\mathbf{x}_1-i\partial_Z\mathbf{x}_1
\\
&-&\mathbf{A}\mathbf{x}_1+i\mathbf{B}\mathbf{x}_0,
\label{eq:third-order}
\end{eqnarray}
with $\mathbf{x}_i = \left[q_i, p_i \right]^\mathrm{T}$ ($i=0,1,2,3$) unknown vectors, and
\begin{align}
\label{eq:matrix-W-A} \mathbf{W} = &\left[ \begin{array}{cc}
-k_0 & \omega_0\mu_L\\
\omega_0\epsilon_L & -k_0\\
\end{array} \right],\,\,\,\,\,\,
\mathbf{A}\mathbf{x}_i = \omega_0\left[
\begin{array}{c}
\beta|p_0|^2p_i\\
\alpha|q_0|^2q_i\\
\end{array} \right],\\[4pt]
\label{eq:matrix-B} \mathbf{B}\mathbf{x}_0 = &\left[
\begin{array}{c}
-\beta\partial_T(|p_0|^2p_0)+i\omega_0\beta(p_0p_1^{\star}+p_0^{\star}p_1)p_0\\
-\alpha\partial_T(|q_0|^2q_0)+i\omega_0\alpha(q_0q_1^{\star}+q_0^{\star}q_1)q_0\\
\end{array} \right],
\end{align}
with $\star$ denoting complex conjugate. To proceed further, we note that
the compatibility conditions required for Eqs. (\ref{eq:zero-order})-(\ref{eq:third-order})
to be solvable, known also as Fredholm alternatives \cite{rpm,kodhas}, are
$\mathbf{L}\mathbf{W}\mathbf{x}_i =0$, where $\mathbf{L}=[1, Z_L]$ is a left eigenvector of
of $\mathbf{W}$, such that $\mathbf{L} \mathbf{W}=\mathbf{0}$, with $Z_L=\sqrt{\mu_L/\epsilon_L}$
being the linear wave-impedance.

The leading-order Eq. (\ref{eq:zero-order}) provides the following results.
First, the solution $\mathbf{x}_0$ of Eq. (\ref{eq:zero-order}) has the form:
\begin{equation}
\label{eq:x0-result} \mathbf{x}_0=\mathbf{R} \phi(Z,T),
\end{equation}
where $\phi(Z,T)$ is an unknown scalar field
and
$\mathbf{R}=[1, Z_L^{-1}]^{\rm T}$ is a right eigenvector of $\mathbf{W}$, such that
$\mathbf{W}\mathbf{R}=\mathbf{0}$. Second, by using the compatibility condition
$\mathbf{L}\mathbf{W}\mathbf{x}_0 =0$ and Eq. (\ref{eq:x0-result}), we obtain
the equation $\mathbf{L}\mathbf{W}\mathbf{R}=0$, which is actually the linear dispersion relation,
\begin{equation}
\label{eq:disp-rel}
k_{0}^{2}=\omega_0^2\epsilon_{L}\mu_{L},
\end{equation}
%
($\epsilon_{L}$ and $\mu_{L}$ are evaluated at $\omega_0$).
Note that Eq. (\ref{eq:disp-rel}) is also obtained by imposing the
nontrivial solution condition ${\rm det}\mathbf{W} = 0$.
Third, the
EM field envelopes are proportional to each other,
i.e., $q_0 = p_0 Z_L$.

%

At $\mathcal{O}(\varepsilon^1)$, the compatibility condition for Eq. (\ref{eq:first-order})
results in $\mathbf{L}\mathbf{W}'\mathbf{R}=0$, written equivalently as:
\begin{equation}
\label{eq:group-veloc}
2k_0 k_0'=\omega_0^2(\epsilon_{L}\mu'_{L}+\epsilon'_{L}\mu_{L})
+2\omega_0\epsilon_{L}\mu_{L}.
\end{equation}
\begin{figure}[t]
\centering
\includegraphics[width=.38\textwidth, height=.2\textheight]{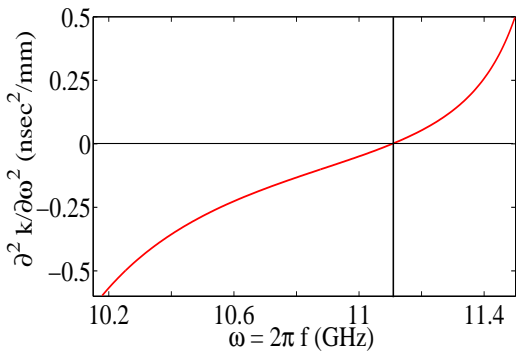}
\caption{ (Color online) The
GVD coefficient $k'' \equiv \partial^2 k/\partial \omega^2$ as
a function of frequency in the
left-handed regime.
}
\label{fig2}
\vskip-0.35cm
\end{figure}

This is actually the definition of the group velocity $v_g = 1/k_{0}'$, as can also be found by differentiating
Eq. (\ref{eq:disp-rel}) with respect to $\omega$. Furthermore, using Eq. (\ref{eq:x0-result}),
Eq. (\ref{eq:first-order}) suggests that the unknown vector $\mathbf{x}_1$ has the form,
\begin{equation}
\label{eq:x1-result}
\mathbf{x}_1=i\mathbf{R}'\partial_T\phi(Z,T)+\mathbf{R}\psi(Z,T),
\end{equation}
where $\psi(Z,T)$ is an unknown scalar field.

Next, at order $\mathcal{O}(\varepsilon^2)$, the compatibility condition for
Eq. (\ref{eq:second-order}), combined with Eqs. (\ref{eq:x0-result}) and (\ref{eq:x1-result}), yields
the following NLS equation,
\begin{equation}
\label{eq:NLS}
i\partial_Z\phi -\frac{1}{2}k_0''
\partial_T^2\phi+\gamma|\phi|^2\phi=0,
\end{equation}
where $k_0''$ is the group-velocity dispersion (GVD) coefficient, as can be evaluated
by differentiating $k_0'$ in Eq. (\ref{eq:group-veloc}), and 
%
$\gamma= (\omega_0^2 / 2k_0)(\epsilon_0 \alpha\mu_L+ \mu_0 \beta\epsilon_LZ_L^{-2})$.
%
Note that once $\phi$ is obtained from the NLS Eq. (\ref{eq:NLS}), the
EM field envelopes
are determined as $q_0 = \phi$ and $p_0 = Z_L^{-1} \phi$ [see Eq. (\ref{eq:x0-result})],
similarly to the case of a linear medium.

Finally, to order $\mathcal{O}(\varepsilon^3)$, we use the compatibility condition for
Eq. (\ref{eq:third-order}), as well as Eqs. (\ref{eq:second-order}), (\ref{eq:x0-result}) and (\ref{eq:x1-result}),
and obtain a NLS equation, incorporating higher-order dispersive and nonlinear terms.
This equation describes the evolution of $\psi$, and yet contains $\phi$, which in turn obeys Eq.
(\ref{eq:NLS}). Instead of considering this system of two equations, we follow \cite{potasek,kodhas,hasbook}
and introduce a new combined function $\Phi=\phi+\varepsilon \psi$. This way, combining the NLS equations obtained at orders
$\mathcal{O}(\varepsilon^2)$ and $\mathcal{O}(\varepsilon^3)$, we find that $\Phi$ obeys the HNLS equation:
\begin{align}
\nonumber i\partial_Z\Phi &-\frac{1}{2}k_0''
\partial_T^2\Phi+\gamma|\Phi|^2\Phi\\
\label{eq:HNLS}
&=i\varepsilon\left[\frac{1}{6}k_0'''\partial_T^3\Phi-\frac{\gamma}{\omega_0}\partial_T(|\Phi|^2\Phi)\right].
\end{align}
%

For $\varepsilon=0$, the HNLS Eq. (\ref{eq:HNLS}) is reduced to the NLS Eq. (\ref{eq:NLS}),
while for $\varepsilon \ne 0$ generalizes the higher-order NLS equation
describing ultra-short pulse propagation in optical fibers \cite{potasek,kodhas,potasek2,hasbook}
(where 
dispersion and nonlinearity appear solely in the 
dielectric properties).
As in the 
NLS Eq. (\ref{eq:NLS}),
Eq. (\ref{eq:HNLS}) provides the field $\Phi$
which, in turn, determines the
EM fields at order $\mathcal{O}(\varepsilon^3)$
as $q_0+\varepsilon q_1 = \Phi$ and $p_0+\varepsilon p_1 = Z_L^{-1} \Phi$ [see
Eqs. (\ref{eq:x0-result}), (\ref{eq:x1-result})]. Finally, we stress that the NLS Eq. (\ref{eq:NLS}),
or the HNLS Eq. (\ref{eq:HNLS}), can be used in the LH (RH) regime, taking
$k_0$, $\epsilon_L$, and $\mu_L$ negative (positive) as per the discussion above.

\begin{table}
\begin{center}
\caption{Conditions for the formation of bright or dark solitons (BS or DS)
for the NLS Eq. (\ref{eq:NLS-scaled}).}
\vskip+0.2cm
\begin{tabular}{|c|c|c|c|}
\hline
$\,$ &  $\,$ & $s=+1$ & $s=-1$\\
\hline
$\sigma=+1$ & $\alpha>0$ & DS & BS \\[8pt]
\hline
$\sigma=-1$ & $\alpha<0$,\,\,$|\frac{\alpha}{\beta}|>\frac{Z_0^2}{Z_L^4}$ & BS & DS \\[8pt]
\hline
$\sigma=+1$ & $\alpha<0$,\,\,$|\frac{\alpha}{\beta}|<\frac{Z_0^2}{Z_L^4}$ & DS & BS \\[8pt]
\hline
\end{tabular}
\end{center}
\label{t1}
\end{table}

Let us now analyze Eqs. (\ref{eq:NLS}) and (\ref{eq:HNLS}) in more detail. First,
measuring length, time, and the field intensity $|\phi|^2$ in units of
the dispersion length $L_D=t_0^2 / |k_0''|$, initial pulse width $t_0$, and $L_D/|\gamma|$, respectively,
we reduce the NLS Eq. (\ref{eq:NLS}) to the following dimensionless form:
\begin{equation}
\label{eq:NLS-scaled} i\partial_{Z}\phi -\frac{s}{2}
\partial_{T}^2 \phi +\sigma |\phi|^2 \phi=0,
\end{equation}
where $s={\rm sign}(k_0'')$ and $\sigma = {\rm sign}(\gamma)$. The NLS Eq. (\ref{eq:NLS-scaled})
admits bright (dark) soliton solutions for $s \sigma =-1$ ($s \sigma =+1$).
As is shown in  Fig. \ref{fig2}, for our choice of parameters,
$s=+1$ (i.e., $k_{0}''>0$) for $2\pi \times 1.76 < \omega <2\pi \times 1.87$ GHz, while
$s=-1$ (i.e., $k_{0}''<0$) for $2\pi \times 1.45 < \omega <2\pi \times 1.76$ GHz
in the
LH regime.
Moreover, since $\beta>0$, we have $\sigma=+1$ either for a focusing dielectric,
$\alpha>0$,
or for a defocusing dielectric, $\alpha<0$, with $|\alpha/\beta| < Z_0^2/Z_L^4$
($Z_0 = \sqrt{\mu_0/\epsilon_0}$
is the vacuum wave-impedance). Hence, for $\sigma=+1$, bright (dark) solitons occur in the anomalous (normal) dispersion regimes,
i.e., for $k_{0}''<0$ ($k_{0}''>0$), respectively. On the other hand, $\sigma=-1$ for
$\alpha<0$, with $|\alpha/\beta| > Z_0^2/Z_L^4$ and,
bright (dark) solitons occur in the normal (anomalous) dispersion regimes. The above results are summarized in Table I.
%
Note that the 
presence of dispersion and nonlinearity 
in the {\it magnetic} response of the
metamaterial 
allows for
bright (dark) solitons in the anomalous (normal) dispersion regimes for {\it defocusing} dielectrics
(see third line of Table I).

Next, we consider the HNLS Eq. (\ref{eq:HNLS}) which, by using the same dimensionless units as
before, is expressed as,
\begin{equation}
i\partial_{Z} \Phi -\frac{s}{2}\partial_{T}^2 \Phi +\sigma|\Phi|^2 \Phi =
i \delta_1 \partial_{T}^3 \Phi -i \sigma \delta_2\partial_{T}(|\Phi|^2 \Phi),
\label{eq:partic-higher-NLS}
\end{equation}
where $\delta_1=\varepsilon k_0'''/(6t_0|k_0''|)$, and 
$\delta_2= \varepsilon /(\omega_0 t_0)$.
%
%
%
Equation (\ref{eq:partic-higher-NLS}) can be used to predict {\it ultra-short
solitons} in nonlinear
LH metamaterials as follows.  
Following Ref. \cite{hizfrapol}, 
we seek travelling-wave solutions
of Eq. (\ref{eq:partic-higher-NLS}) of the form, 
%
\begin{equation}
\Phi(Z,T)= U(\eta) \exp[i(K Z -\Omega T)],
\label{eq:gauge-transformation}
\end{equation}
where $U(\eta)$ is the unknown envelope function (assumed to be
real), $\eta = T - \Lambda Z$, and the real parameters $\Lambda$,
$K$ and $\Omega$ denote, respectively, the inverse velocity,
wavenumber and frequency of the travelling wave. Substituting Eq.
(\ref{eq:gauge-transformation}) into Eq. (\ref{eq:partic-higher-NLS}), 
the real and imaginary parts of the resulting equation respectively read: 
\begin{align}
&\ddot{U}+\frac{K-\frac{s}{2}\Omega^2-\delta_1\Omega^3}{\frac{s}{2}+3\delta_1\Omega}U
-\frac{\sigma(1+\delta_2\Omega)}{{\frac{s}{2}+3\delta_1\Omega}}U^3=0,
\label{ode1} \\
&\delta_1 \dddot{U}+(\Lambda-s\Omega-3\delta_1\Omega^2)\dot{U}-3\sigma\delta_2 U^2 \dot{U}=0,
\label{ode2}
\end{align}
where overdots denote differentiations with respect to $\eta$. Notice that in the case of 
$\delta_1 =\delta_2 =0$, Eq. (\ref{ode2}) is automatically satisfied if $\Lambda = s \Omega$ and the profile 
of ``long'' soliton pulses [governed by Eq. (\ref{eq:NLS-scaled})] is determined by Eq. (\ref{ode1}). 
On the other hand, for ultra-short solitons (corresponding to $\delta_1 \ne 0$, $\delta_2 \ne 0$), 
the system of Eqs. (\ref{ode1}) and (\ref{ode2}) is consistent if the following conditions hold:
\begin{align}
\label{eq:kappa} 
&\frac{K -\frac{s}{2} \Omega^2 - \delta_1
\Omega^3}{\frac{s}{2}+3\delta_1\Omega}=\frac{\Lambda-s\Omega-3\delta_1\Omega^2}{\delta_1} \equiv \kappa,\\
\label{eq:nu}
&-\frac{\sigma\delta_2}{\delta_1}=-\frac{\sigma(1+\delta_2\Omega)}{\frac{s}{2}+3\delta_1\Omega} \equiv \nu,
\end{align}
where $\kappa$ and $\nu$ are nonzero constants. In such a case, 
Eqs. (\ref{ode1}) and (\ref{ode2}) are equivalent to 
the following equation of motion of the unforced and
undamped Duffing oscillator,
\begin{equation}
\ddot{U} + \kappa U +\nu U^3=0. \label{eq:Duffing}
\end{equation}
For $\kappa \nu <0$, Eq. (\ref{eq:Duffing}) possesses two exponentially
localized solutions (as special cases of its general elliptic
function solutions), corresponding to the separatrices in the $(U, \dot{U})$ phase-plane.  
These solutions have the form of a hyperbolic secant (tangent) for
$\kappa<0$ and $\nu>0$ ($\kappa>0$ and $\nu<0$), thus
corresponding to the bright, $U_{\rm BS}$ (dark, $U_{\rm DS}$) solitons of Eq.
(\ref{eq:partic-higher-NLS}):
%
\begin{eqnarray}
U_{\rm BS}(\eta) &=& \left(2|\kappa|/\nu\right)^{1/2}
\mathrm{sech}(\sqrt{|\kappa|}\eta),
\label{bs} \, \qquad \\
U_{\rm DS}(\eta) &=& \left(2\kappa/|\nu|\right)^{1/2}
\mathrm{tanh}(\sqrt{\kappa/2}\eta). \label{ds}
\end{eqnarray}
%
These are {\it ultra-short} solitons of the HNLS Eq.
(\ref{eq:partic-higher-NLS}), valid even for $\varepsilon =
\mathcal{O}(1)$: since both coefficients $\delta_1$, $\delta_2$ of Eq.
(\ref{eq:partic-higher-NLS}) scale as $\varepsilon (\omega_0
t_0)^{-1}$, it is clear that for $\omega_0 t_0 = \mathcal{O}(1)$,
or for soliton widths $t_0 \sim \omega_0^{-1}$, the higher-order
terms can safely be neglected and soliton propagation is governed
by Eq. (\ref{eq:NLS-scaled}). On the other hand, if $\omega_0 t_0 =
\mathcal{O}(\varepsilon)$, the higher-order terms become important
and solitons governed by the HNLS Eq. (\ref{eq:partic-higher-NLS})
are {\it ultra-short}, of a width $t_0 \sim \varepsilon
\omega_0^{-1}$. We stress that these solitons
are approximate solutions of Maxwell's equations, satisfying
Faraday's and Amp\'{e}re's Laws in Eqs. (\ref{eq:Faraday-Ampere-laws})
up to order $\mathcal{O}(\varepsilon^3)$.

Finally, as concerns the condition for bright or dark soliton
formation, namely $\kappa \nu <0$, we note that $\kappa$
depends on the free parameters $K$ and $\Omega$ (and, thus, can be tuned
on demand), while the parameter $\nu$ has the opposite sign from $\sigma$
(since $\delta_2>0$, while ${\rm sign}(\delta_1)={\rm sign}(k_0''') = +1$ -- see Fig. \ref{fig2}).
This means that bright solitons are formed for $\kappa<0$ and $\sigma =-1$ (i.e., $\alpha<0$ with $|\alpha/\beta| > Z_0^2/Z_L^4$),
while dark ones are formed for $\kappa>0$ and $\sigma =+1$ (i.e., $\alpha>0$, or $\alpha<0$ with $|\alpha/\beta| < Z_0^2/Z_L^4$).

In conclusion, we used the reductive perturbation method to 
derive from Maxwell's equations a HNLS equation describing 
pulse propagation in nonlinear metamaterials. 
We studied 
the pertinent dispersive and nonlinear effects, 
found necessary conditions
for the formation of
bright or dark ultra-short solitons, as well as
approximate
analytical expressions for these solutions. 
Future research may include a systematic study of the stability and dynamics
of the ultra-short solitons, both in the framework of the HNLS equation and, perhaps more
importantly, in the context of Maxwell's equations.



\begin{thebibliography}{99}


\bibitem{exp} D. R. Smith {\it et al.},
Phys. Rev. Lett. {\bf 84}, 4184 (2000);
D. R. Smith and N. Kroll,
Phys. Rev. Lett. {\bf 85}, 2933 (2000);
A. Shelby, D. R. Smith, and S. Schultz,
Science {\bf 292}, 77 (2001).

\bibitem{reviews} D. R. Smith, J. B. Pendry, M. C. K. Wiltshire,
Science {\bf 305}, 788 (2004);
C. M. Soukoulis, M. Kafesaki, and E.N. Economou, Adv. Materials {\bf 18}, 1941 (2006);
%
G. V. Eleftheriades and K. G. Balmain (eds.) {\it Negative-Refraction Metamaterials. Fundamental Principles
and Applications} (John Wiley, New Jersey, 2005).

\bibitem{zharov} A. A. Zharov, I. V. Shadrivov, and Yu. S. Kivshar,
Phys. Rev. Lett. \textbf{91}, 037401 (2003).

\bibitem{agranovich} V. M. Agranovich {\it et al.},
Phys. Rev. B \textbf{69}, 165112 (2004).

\bibitem{shadrivov} I. V. Shadrivov {\it et al.},
Radio Sci. \textbf{40}, RS3S90 (2005).

\bibitem{lazarides-tsironis} N. Lazarides, and G. P. Tsironis,
Phys. Rev. E \textbf{71}, 036614 (2005).

\bibitem{manakov} S. V. Manakov, Zh. Eksp. Teor. Fiz. \textbf{65}, 505
(1973) [Sov. Phys. JETP \textbf{38}, 248 (1974)].

\bibitem{kourakis} I. Kourakis, and P. K. Shukla, Phys. Rev. E \textbf{72}, 016626 (2005).

\bibitem{shukla} M. Marklund {\it et al.},
Phys. Lett. A {\bf 341}, 231 (2005).

\bibitem{scalora} M. Scalora {\it et al.},
Phys. Rev. Lett. \textbf{95}, 013902 (2005).


\bibitem{wen-pre} S. C. Wen {\it et al.},
Phys. Rev. E \textbf{73}, 036617 (2006).

\bibitem{shukla2} M. Marklund, P. K. Shukla, and L. Stenflo, Phys. Rev. E {\bf 73}, 037601 (2006).

\bibitem{wen-pra} S. C. Wen {\it et al.},
Phys. Rev. A \textbf{75}, 033815 (2007).

\bibitem{rpm} T. Taniuti, Prog. Theor. Phys. Suppl. {\bf 55}, 1 (1974);
%
H. Leblond, J. Phys. B
{\bf 41}, 043001 (2008).


\bibitem{potasek} Y. Kodama, J. Stat. Phys. {\bf 39}, 597 (1985).

\bibitem{kodhas} Y. Kodama and A. Hasegawa, IEEE J. Quantum Electron. {\bf 23}, 510 (1987).

\bibitem{potasek2} M. J. Potasek, J. Appl. Phys. {\bf 65}, 941 (1989).

\bibitem{hasbook} A. Hasegawa and Y. Kodama, {\it Solitons in Optical Communications}
(Clarendon Press, Oxford, 1995).


\bibitem{yskss} I. V. Shadrivov and Y. S. Kivshar, J. Opt. A: Pure Appl. Opt. {\bf 7}, S68 (2005).


\bibitem{hizfrapol} K. Hizanidis, D. J. Frantzeskakis, and C. Polymilis, J. Phys. A: Math. Gen. {\bf 29}, 7687 (1996).















\end{thebibliography}
\end{document}